\documentclass[12pt,a4paper]{article}
\usepackage[latin9]{inputenc}
\usepackage{times}
\usepackage{a4wide}
\usepackage{amssymb}
\usepackage{amsmath}
\usepackage[notref,notcite]{showkeys} 
\usepackage{ifpdf}
\ifpdf
\usepackage[pdftex]{graphicx}
\usepackage[pdftex,unicode,implicit]{hyperref}
\hypersetup{%
  pdftitle    = {A Simple Derivation of Supersymmetric Extremal BH Attractors},
  pdfkeywords = {supersymmetry, black holes, attractors},
  pdfauthor   = {T. Ortín},
  pdfcreator  = {pdf\LaTeXe\ with package \flqq hyperref\frqq},
  pdfproducer = {pdf\LaTeXe\ with package \flqq hyperref\frqq},
  pdfpagemode = None,  
  pdffitwindow= true,  
  unicode     = true,
  plainpages  = false,
  colorlinks  = true,  
  citecolor   = blue,
  urlcolor    = red,
  linkcolor   = black
}
\newcommand{\hepth}[1]{({\tt \href{http://www.arXiv.org/abs/hep-th/#1}{hep-th/#1}})}

\newcommand{\arxiv}[1]{{\tt \href{http://www.arXiv.org/abs/#1}{arXiv:#1}}}
\else
  \usepackage[dvips]{graphicx}
  \usepackage[unicode,implicit]{hyperref}
  \newcommand{\hepth}[1]{{\tt hep-th/#1}}

  \newcommand{\arxiv}[1]{{\tt arXiv:#1}}
\fi
\makeatletter
\@addtoreset{equation}{section}
\makeatother

\begin{document}
\begin{flushright}
\small
IFT-UAM/CSIC-10-40\\
March $16^{\rm th}$, $2011$
\normalsize
\end{flushright}
\begin{center}
\vspace{2cm}
{\Large {\bf A Simple Derivation of Supersymmetric}}\\[.5cm]
{\Large {\bf Extremal Black-Hole Attractors}}
\vspace{2cm}

{\sl\large Tom{\'a}s Ort\'{\i}n}
\footnote{E-mail: {\tt Tomas.Ortin@csic.es}}

\vspace{1cm}

{\it Instituto de F\'{\i}sica Te\'orica UAM/CSIC\\
C/ Nicol\'as Cabrera, 13-15,  C.U.~Cantoblanco,  E-28049-Madrid, Spain}\\

\vspace{2cm}


{\bf Abstract}

\end{center}

\begin{quotation}\small
  We present a simple and yet rigorous derivation of the flow equations for
  the supersymmetric black-hole solutions of all 4-dimensional supergravities
  based on the recently found general form of all those solutions.
\end{quotation}

\newpage

\pagestyle{plain}




\section*{Introduction}

The discovery of the attractor mechanism that drives the scalar fields of
supersymmetric extremal black holes to take values that only depend on the
electric and magnetic charges on the event horizon
\cite{Ferrara:1995ih,Strominger:1996kf,Ferrara:1996dd,Ferrara:1996um,Ferrara:1997tw,Ferrara:2006em}
has undoubtedly been one of the mean breakthroughs of black-hole physics in
the recent years. Long after the discovery of the existence of extremal but
non-supersymmetric black-hole solutions
\cite{Khuri:1995xq,Ortin:1996bz,Dabholkar:1997rk} it was realized that there
is an attractor mechanism at work in those black holes as well.
\cite{Sen:2005wa,Goldstein:2005hq,Tripathy:2005qp,Kallosh:2006bt}. Since the
existence of the attractor mechanism in supersymmetric black holes is related
to the existence of flow (first-order) equations for the metric function and
scalar fields that follow from the Killing spinor equations\footnote{See,
  e.g.~Ref.~\cite{Bellorin:2006xr} for a derivation of the flow equations in
  $N=2,d=4$ supergravity along these lines.}, it was natural to search for
flow equations driving the metric function and scalar fields of extremal
non-BPS black holes to their attractor values. Those equations, which depend
on a ``superpotential'' function which coincides with the central charge in
the BPS case were found in Ref.~\cite{Ceresole:2007wx} for $N=2,d=4$
supergravity and in Ref.~\cite{Ceresole:2007wx} for $N>2,d=4$ theories. These
developments were based in the approach pioneered in Ref.\cite{}. Further
extensions of these results to non-supersymmetric cases and singular
black-hole-type solutions (``small back holes'') can be found in
Refs.~\cite{Bossard:2009we,Ceresole:2009vp,Andrianopoli:2009je,Ceresole:2010nm}.

In this paper we are going to present a simple derivation of those flow
equations in all 4-dimensional, ungauged, supergravities which does not make
explicit use of supersymmetry and may be valid for extremal non-supersymmetric
black holes and other solutions of those theories.

We start by deriving in Section~\ref{sec-d5} the black-hole flow equations for
$N=2,d=5$ supergravity coupled to vector supermultiplets as a particularly
simple example. In Section~\ref{sec-n2d4} we work out the well-known case of
$N=2,d=4$ supergravity coupled to vector supermultiplets and in
Section~\ref{sec-generalnd4} we generalize our results to general $N,d=4$
supergravity.


\section{$N=2,d=5$}
\label{sec-d5}

The $N=2,d=5$ vector supermultiplets contain one 1-form $A^{x}{}_{\mu}$ and
one real scalar $\phi^{x}$ ($x,y,z=1,\cdots, n$ where $n$ is the number of
vector multiplets). The $n$ matter 1-forms are combined with the graviphoton
$A^{0}{}_{\mu}$ into $A^{I}{}_{\mu}$ and the $n$ scalars are described by
$\bar{n}=n+1$ real functions $h_{I}(\phi)$, ($I,J,K=0,1,\cdots n$). These are
subject to the constraints

\begin{equation}
\label{eq:constraintshI}
h^{I}h_{I}=1\, ,
\hspace{1cm}
h^{I}dh_{I}=h_{I}dh^{I}=0\, . 
\end{equation}

\noindent
The metric is the scalar manifold $g_{xy}$ is given by 

\begin{equation}
\label{eq:defgxy}
h^{I}{}_{x}h_{Iy}=g_{xy}\, ,
\hspace{1cm}
h_{Iy}\equiv -\sqrt{3}\partial_{y}h_{I}\, ,
\hspace{1cm}
h^{I}{}_{x}=\sqrt{3}\partial_{x}h_{I}\, .
\end{equation}

It is well known that the static, spherically-symmetric supersymmetric
solutions of these theories are such that the quotients $h_{I}/f$, where $f$
is related to the spacetime metric by $f^{2}=g_{tt}$, are harmonic functions
in Euclidean $\mathbb{R}^{4}$ \cite{Gauntlett:2002nw,Gauntlett:2004qy}. We can
write these functions as linear functions of an appropriate coordinate $\tau$

\begin{equation}
\label{eq:hIf}
h_{I}/f \equiv l_{I} -q_{I}\tau\, ,
\end{equation}

\noindent
where the constants $q_{I}$ are the electric charges. We are going to assume
that we have a field configuration of the above form for some coordinate
$\tau$, not necessarily supersymmetric, not necessarily satisfying the
equations of motion and not necessarily being a black hole, although
supersymmetric black holes would be the prime example to which one can apply
the following results.

The central charge in these theories is given by

\begin{equation}
\mathcal{Z}[\phi,q]\equiv h^{I}(\phi)q_{I}\, .
\end{equation}

We are going to find the flow equations obeyed by $f(\tau)$ and the scalar
fields $\phi^{x}(\tau)$ using just basic relations of real special
geometry. First, using Eqs.~(\ref{eq:constraintshI}) we write the differential
of $f^{-1}$ as

\begin{equation}
df^{-1} = d(h^{I} h_{I}/f) = h^{I} d (h_{I}/f)\, ,
\end{equation}

\noindent
from which we get the first component of the flow equations

\begin{equation}
\label{eq:d5floweq1}
\frac{df^{-1}}{d\tau} = \mathcal{Z}[\phi(\tau),q]\, .
\end{equation}

Using now the same constraints plus the definition Eq.~(\ref{eq:defgxy}) we
can write for the differential of $\phi^{x}$

\begin{equation}
d\phi^{x} = h^{Ix}h_{Iy}d\phi^{y} 
= -\sqrt{3}h^{Ix} d h_{I} =  -\sqrt{3}h^{Ix} d (f h_{I}/f)
= -\sqrt{3} f h^{Ix} d (h_{I}/f)\, ,
\end{equation}

\noindent
from which we get the remaining $n$ components of the flow equations:

\begin{equation}
\label{eq:d5floweq2}
\frac{d\phi^{x}}{d\tau} = f g^{xy}\partial_{y}\mathcal{Z}[\phi(\tau),q] \, .
\end{equation}

The variables $\phi^{x}(\tau)$ and the solutions will be attracted to the
fixed points $\phi^{x}_{\rm fixed}$ at which the r.h.s.~vanishes, i.e.~where
the attractor equations

\begin{equation}
  \partial_{y}\mathcal{Z}[\phi^{x}_{\rm fixed},q]=0\, ,  
\end{equation}

\noindent
are satisfied. The solutions of these equations give $\phi^{x}_{\rm fixed}$ as
functions of the electric charges $q_{I}$ and at the point $\tau=\tau_{\rm
  fixed}$, $\phi^{x}$ takes the value $\phi^{x}_{\rm fixed}(q)$, independently
of the constants $l_{I}$. Furthermore, at the attractor point

\begin{equation}
\left. \frac{df^{-1}}{d\tau}\right|_{\tau=\tau_{\rm fixed}} 
= \mathcal{Z}[\phi_{\rm fixed}(q),q]\equiv \mathcal{Z}_{\rm fixed}(q)\, .
\end{equation}

The derivation of the flow equations (\ref{eq:d5floweq1}),(\ref{eq:d5floweq2})
that we have presented and the properties that follow (the attractor
mechanism) holds for any field configuration of the form Eq.~(\ref{eq:hIf}),
irrespectively of the meaning of the function $f$ or the coordinate $\tau$ and
of the physical properties of the configuration.  If the field configuration
describes a 5-dimensional supersymmetric black-hole solution, then one can
show that there is an event horizon at $\tau_{\rm fixed}$ and the attractor
mechanism relates the central charge to the black-hole entropy
\cite{Ferrara:1996um}. The general 5-dimensional flow equations for $N=2$
theories have been derived in Ref.~\cite{Lopes Cardoso:2007ky}, in
Ref.~\cite{Mohaupt:2009iq} using timelike dimensional reduction techniques and
in Ref.~\cite{Mohaupt:2009iq}.


\section{$N=2,d=4$}
\label{sec-n2d4}

Let us now consider $N=2,d=4$ supergravity coupled to $n$ vector
supermultiplets.  Each of them contains a 1-form $A^{i}{}_{\mu}$ and one
complex scalar $Z^{i}$ $i=1,\cdots, n$. The $Z^{i}$ parametrize a special
K\"ahler manifold. The $n$ matter 1-forms are combined with the graviphoton
into $A^{\Lambda}{}_{\mu}$ ($\Lambda,\Sigma =0,1,\cdots,n$) while the $n$
complex scalars are combined into the $2\bar{n}=2(n+1)$ components of the
symplectic section
$\mathcal{V}\equiv(\mathcal{L}^{\Lambda},\mathcal{M}_{\Lambda})$. These are
subject to the constraints

\begin{equation}
\label{eq:constraintsVN=2}
  i\langle\, \mathcal{V}\mid \mathcal{V}^{*}\, \rangle \equiv 
  i(\mathcal{L}^{*\, \Lambda}\mathcal{M}_{\Lambda} 
  -\mathcal{M}^{*}_{\Lambda}\mathcal{L}_{\Lambda} ) =1\, ,
  \hspace{1cm} 
  \langle\, \mathcal{D}_{i}\mathcal{V}\mid \mathcal{V}^{*}\, \rangle = 0\, ,  
\end{equation}

\noindent
where $\mathcal{D}_{i}$ is the K\"ahler-covariant derivative and $\mathcal{V}$
has K\"ahler weight $1$.

The supersymmetric black-hole solutions of these theories
\cite{Ferrara:1995ih,Strominger:1996kf,Sabra:1997kq,Sabra:1997dh,Behrndt:1997ny,LopesCardoso:2000qm,Meessen:2006tu}
are such that the components of the real symplectic vector $\mathcal{I}\equiv
\Im{\rm m}(\mathcal{V}/X)$, where $X$ is a K\"ahler-weight\footnote{The
  function $X$ appears naturally in the spinor-bilinear method
  \cite{Meessen:2006tu} and plays a role analogous to that of the function $f$
  in the $N=2,d=5$ case. Furthermore, since $X$ has the same K\"ahler weight
  as $\mathcal{V}$, the quotient is K\"ahler gauge-independent. This
  independence is necessary for the prescription to construct the most general
  black-hole solutions to be consistent.}  $1$ complex function related to the
spacetime metric by $|X|^{-2}=2g_{tt}$, are given again by linear functions of
some coordinate $\tau$

\begin{equation}
\mathcal{I} \equiv \mathcal{I}_{0}  -\tfrac{1}{\sqrt{2}}\mathcal{Q}\tau\, ,
\end{equation}

\noindent
where $\mathcal{I}_{0}$ and $\mathcal{Q}$ are constant symplectic
vectors\footnote{The factor $1/\sqrt{2}$, necessary for a correct
  normalization for the charges, was omitted in
  Ref.~\cite{Bellorin:2006xr}.}. The components of $\mathcal{Q}$,
$(p^{\Lambda},q_{\Lambda})$, are the magnetic and electric charges of the
solution.

We are going to assume that we have a field configuration of the above form
for some coordinate $\tau$, not necessarily supersymmetric, not necessarily
satisfying the equations of motion and not necessarily being a black hole, and
we are going to find flow equations for $X(\tau)$ and the complex scalar
fields $Z^{i}(\tau)$ using basic relations of special geometry.

Let us define the central charge

\begin{equation}
  \mathcal{Z}[Z(\tau),\mathcal{Q}]\equiv 
\langle\, \mathcal{V}\mid \mathcal{Q}\, \rangle = 
p^{\Lambda}\mathcal{M}_{\Lambda} - q_{\Lambda}\mathcal{L}^{\Lambda}\, .
\end{equation}

Since $\mathcal{V}/X$ has zero K\"ahler weight, using Eqs.~(\ref{eq:constraintsVN=2})

\begin{equation}
\mathcal{D}X^{-1} = 
i\langle\, \mathcal{V}\mid \mathcal{V}^{*}\, \rangle\, 
\mathcal{D}X^{-1} = 
\langle\, \mathcal{D}(\mathcal{V}/X)\mid \mathcal{V}^{*}\, \rangle  
= \langle\, d(\mathcal{V}/X)\mid \mathcal{V}^{*}\, \rangle\, .  
\end{equation}

\noindent
We now need to use a less trivial property, proved in an appendix of
Ref.~\cite{Bellorin:2006xr} using the homogeneity of the prepotential

\begin{equation}
\label{eq:property}
\langle\, d(\mathcal{V}/X)\mid \mathcal{V}^{*}/X^{*}\, \rangle =
2i \langle\, d\mathcal{I}\mid \mathcal{V}^{*}/X^{*}\, \rangle\, ,
\end{equation}

\noindent
which leads us to 

\begin{equation}
\label{eq:N2d4floweq1}
\mathcal{D}X^{-1} = 
2 \langle\,\mathcal{V}^{*} \mid  d\mathcal{I}\, \rangle\, ,
\end{equation}

\noindent
from which we get the first component of the flow equations

\begin{equation}
\mathcal{D}_{\tau} X^{-1} = -\sqrt{2} \mathcal{Z}^{*}[Z(\tau),\mathcal{Q}]\, .  
\end{equation}

Using the property $-i\langle\, \mathcal{D}_{i}\mathcal{V}\mid
\mathcal{D}_{i^{*}}\mathcal{V}^{*}\, \rangle = \mathcal{G}_{ii^{*}}$
and the previous ones

\begin{equation}
  \begin{array}{rcl}
dZ^{i} & = & i\mathcal{G}^{ij^{*}}\langle\, \mathcal{D}_{j^{*}}\mathcal{V}^{*} \mid
\mathcal{D}_{j}\mathcal{V}\, \rangle   dZ^{j}=
iX\mathcal{G}^{ij^{*}}\langle\, \mathcal{D}_{j^{*}}\mathcal{V}^{*} \mid
\mathcal{D}_{j}(\mathcal{V}/X)\, \rangle   dZ^{j}\\
& & \\
& = & iX\mathcal{G}^{ij^{*}}\langle\, \mathcal{D}_{j^{*}}\mathcal{V}^{*} \mid
\partial_{j}(\mathcal{V}/X)\, \rangle   dZ^{j}=
iX\mathcal{G}^{ij^{*}}\langle\, \mathcal{D}_{j^{*}}\mathcal{V}^{*} \mid
d(\mathcal{V}/X)\, \rangle \\
& & \\
& = &  iX\mathcal{G}^{ij^{*}} \mathcal{D}_{j^{*}}\langle\,\mathcal{V}^{*} \mid
d(\mathcal{V}/X)\, \rangle = -2X\mathcal{G}^{ij^{*}} \mathcal{D}_{j^{*}}
\langle\,\mathcal{V}^{*} \mid d\mathcal{I}\, \rangle\, ,\\
\end{array}
\end{equation}

\noindent
from which we get the remaining components of the flow equations\footnote{For
  a derivation of these equations for black holes, from the Killing spinor
  equations, see Ref.~\cite{Bellorin:2006xr}.}

\begin{equation}
\label{eq:N2d4floweq2}
\frac{dZ^{i}}{d\tau} = \sqrt{2}X\mathcal{G}^{ij^{*}}
\mathcal{D}_{j^{*}}\mathcal{Z}[Z(\tau),\mathcal{Q}]\, .
\end{equation}

The first component of the flow equations, for the black-hole case, is
customarily written in terms of the component $g_{tt}$ of the metric (or the
function $U=\tfrac{1}{2}\log{g_{tt}}$)
\cite{Ferrara:1995ih,Strominger:1996kf,Ferrara:1996dd,Ferrara:1997tw}. Those
expressions can be obtained from Eq.~(\ref{eq:N2d4floweq1}), which is more
general.


\section{Arbitrary $N\geq 2, d=4$}
\label{sec-generalnd4}

In Ref.~\cite{Andrianopoli:1996ve} a formulation of all $N\geq 2, d=4$
supergravities coupled to vector supermultiplets was given that allows to
treat simultaneously all of them\footnote{Some details are $N$-dependent, but
  their treatment or use can be postponed until the end of the analysis.}. This
formulation was recently used in Ref.~\cite{Meessen:2010fh} to determine the
form of all the timelike supersymmetric solutions (including black holes) of
these theories in a unified way. We are going to use this formulation in order
to derive flow equations for the metric function and the scalars of all these
theories.

The scalars of these theories are described by two sets of symplectic vectors:
$\mathcal{V}_{IJ}=\mathcal{V}_{[IJ]},\mathcal{V}_{i}$, where $I,J,K=1,\cdots,
N$ and $i=1,\cdots, n$, 

\begin{equation}\label{eq:symsec}
\mathcal{V}_{IJ}= 
\left(
  \begin{array}{c}
f^{\Lambda}{}_{IJ} \\ h_{\Lambda IJ} \\
  \end{array}
\right)\, ,  
\hspace{1cm}
\mathcal{V}_{i}= 
\left(
  \begin{array}{c}
f^{\Lambda}{}_{i} \\ h_{\Lambda\, i} \\
  \end{array}
\right)\, ,  
\end{equation}

\noindent
$n$ being the number of vector supermultiplets (none
for $N>4$). The theory contains $N(N-1)/2+n$ 1-forms $A^{\Lambda}{}_{\mu}$
$\Lambda=1,\cdots, N(N-1)/2+n$ the first $N(N-1)/2$ of which are the
graviphotons that we could have labeled by $A^{IJ}{}_{\mu}=-A^{JI}{}_{\mu}$
and the rest of which are the matter 1-forms. In all cases, the symplectic
vectors satisfy the constraints

\begin{equation}
  \begin{array}{rcl}
\langle \mathcal{V}_{IJ}\mid\mathcal{V}^{*\, KL}\rangle 
& = &   
-2i\delta^{KL}{}_{IJ}\, , \\
& & \\
\langle \mathcal{V}_{i}\mid\mathcal{V}^{*\, j}\rangle 
&  = &   
-i\delta_{i}{}^{j}\, , \\
\end{array}
\end{equation}

\noindent
with the rest of the symplectic products vanishing. In the $N=2$ case, these
vectors are related to the objects used in the previous section by 

\begin{equation}
\mathcal{V}_{IJ} = \mathcal{V} \varepsilon_{IJ}\, ,
\hspace{1cm}
\mathcal{V}_{i} = \mathcal{D}_{i}\mathcal{V}\, .  
\end{equation}

Using them one can construct the scalar Vielbeine

\begin{equation}
P_{IJKL}= P_{[IJKL]} \equiv -i \langle d
\mathcal{V}_{IJ}\mid\mathcal{V}_{KL}\rangle\, ,
\hspace{1cm}
P_{iIJ} = P_{i[IJ]} \equiv -i \langle d \mathcal{V}_{IJ}\mid\mathcal{V}_{i}\rangle\, .  
\end{equation}

To construct supersymmetric black-hole solutions one must choose first a
time-independent, rank-2 complex antisymmetric matrix $M_{IJ}$ satisfying
$M_{[IJ}M_{K]L} =0$ and

\begin{equation}
M^{[IJ}\mathfrak{D} M^{KL]} = 0\, , 
\end{equation}

\noindent
where $\mathfrak{D}$ s the $U(N)$-covariant derivative. in the $N=2$ case
$M_{IJ}= X\varepsilon_{IJ}$ where $X$ is the complex K\"ahler weight 1
function we introduced in the previous section. $M_{IJ}$ is related to the
spacetime metric by $g_{tt}= |M|^{-2}$ where $|M|^{2} \equiv
M^{PQ}M_{PQ}$. The supersymmetric black-hole solutions are such that the
components of the real symplectic vector

\begin{equation}
\mathcal{I}\equiv \Im {\rm m}\mathcal{V}\equiv 
\Im {\rm m}\left( \mathcal{V}_{IJ}M^{IJ}/|M|^{2}\right)\, ,  
\end{equation}

\noindent
are harmonic functions in Euclidean $\mathbb{R}^{3}$, so they can be written
as linear functions of some coordinate $\tau$

\begin{equation}
\mathcal{I} \equiv \mathcal{I}_{0}  -\tfrac{1}{\sqrt{2}}\mathcal{Q}\tau\, .
\end{equation}

\noindent
where, again $\mathcal{Q}$ is the symplectic vector of all magnetic and
electric charges of the theory.

We are going to show that, for any field configuration of the above form there
are flow equations for the metric function and the scalar Vielbeine.

We  define the central charges

\begin{eqnarray}
\mathcal{Z}_{IJ}[\phi(\tau),\mathcal{Q}] & \equiv & 
 \langle\, \mathcal{V}_{IJ}\mid \mathcal{Q}\, \rangle = 
p^{\Lambda}h_{\Lambda\, IJ} -q_{\Lambda}f^{\Lambda}{}_{IJ}\, ,\\
& & \nonumber \\
\mathcal{Z}_{i}[\phi(\tau),\mathcal{Q}] & \equiv & 
 \langle\, \mathcal{V}_{i}\mid \mathcal{Q}\, \rangle = 
p^{\Lambda}h_{\Lambda\, i} -q_{\Lambda}f^{\Lambda}{}_{i}\, .
\end{eqnarray}

Then, using the above constraints and the definitions of the Vielbeine

\begin{eqnarray}
\label{eq:identity}
\mathfrak{D} \frac{M^{IJ}}{|M|^{2}} & = &   
{\textstyle\frac{i}{2}}\mathfrak{D} \left( \frac{M^{KL}}{|M|^{2}} 
\langle\, \mathcal{V}_{KL} \mid \mathcal{V}^{*\, IJ}\, \rangle \right)
= {\textstyle\frac{i}{2}}\mathfrak{D} 
\langle\, \mathcal{V} \mid \mathcal{V}^{*\, IJ}\, \rangle 
= {\textstyle\frac{i}{2}} 
\langle\, d\mathcal{V} \mid \mathcal{V}^{*\, IJ}\, \rangle \nonumber \\
& & \nonumber \\
& = & 
{\textstyle\frac{i}{2}}
\langle\, d\mathcal{V}^{*} \mid \mathcal{V}^{*\, IJ}\, \rangle
- \langle\, d\mathcal{I} \mid \mathcal{V}^{*\, IJ}\, \rangle
=
{\textstyle\frac{i}{2}}\frac{M_{KL}}{|M|^{2}}
\langle\, d\mathcal{V}^{*\, KL} \mid \mathcal{V}^{*\, IJ}\, \rangle
+\tfrac{1}{\sqrt{2}}\langle\, \mathcal{Q} \mid \mathcal{V}^{*\, IJ}\, \rangle d\tau \nonumber \\
& & \nonumber \\
& = & 
-\tfrac{1}{\sqrt{2}}\mathcal{Z}^{*\, IJ}[\phi(\tau),\mathcal{Q}] d\tau
+{\textstyle\frac{1}{2}}
P^{*\, KLIJ} \frac{M_{KL}}{|M|^{2}}\, .
\end{eqnarray}

\noindent
Using this identity we can compute 

\begin{equation}
\label{eq:compute}
M^{[IJ}\mathfrak{D} \frac{M^{KL]}}{|M|^{2}} =
-\tfrac{1}{\sqrt{2}} M^{[IJ} \mathcal{Z}^{*\, KL]}[\phi(\tau),\mathcal{Q}] d\tau
+{\textstyle\frac{1}{2}}
P^{*\, MN[IJ}\mathcal{J}^{K}{}_{M}\mathcal{J}^{L]}{}_{N}\, ,
\end{equation}

\noindent
where 

\begin{equation}
\mathcal{J}^{I}{}_{J} \equiv 2 |M|^{-2}M^{IK} M_{JK}\, ,
\end{equation}

\noindent
is a rank-2 projector $\mathcal{J}^{2}=\mathcal{J}$,
$\mathcal{J}^{I}{}_{I}=2$. $M_{IJ}$ and $\mathcal{J}$ project over and $N=2$
subspace of the theory and induces a decomposition of all objects into $N=2$
representations. In particular, the projected Vielbeine $P^{*\,
  MN[IJ}\mathcal{J}^{K}{}_{M}\mathcal{J}^{L]}{}_{N}$ and $P_{i\, KL}
\mathcal{J}^{K}{}_{I}\mathcal{J}^{L}{}_{J}$ would correspond to scalars in
$N=2$ vector supermultiplets. The remaining components of the scalar Vielbein
would correspond to $N=2$ hyperscalar which do not have any attractor behavior
and do not allow for regular black hole solutions when they are excited
\cite{Huebscher:2006mr} and therefore will not be considered any
further\footnote{See also
  \cite{Andrianopoli:1997pn,Ferrara:2007tu,Andrianopoli:2010bj}.}.

Since the l.h.s.~of Eq.~(\ref{eq:compute}) vanishes, we get the flow equation
for scalars ($N=4,6,8$)

\begin{equation}
  P^{*\, MN[IJ}\mathcal{J}^{K}{}_{M}\mathcal{J}^{L]}{}_{N}
  = \sqrt{2}
  M^{[IJ} \mathcal{Z}^{*\, KL]}[\phi(\tau),\mathcal{Q}] d\tau\, .
\end{equation}

We can also compute from Eq.~(\ref{eq:identity}) 

\begin{equation}
d|M|^{-2}
=   
\frac{M_{IJ}}{|M|^{2}}\mathfrak{D} \frac{M^{IJ}}{|M|^{2}} 
+
\frac{M^{IJ}}{|M|^{2}}\mathfrak{D} \frac{M_{IJ}}{|M|^{2}} 
=
-\tfrac{1}{\sqrt{2}}
\left[\frac{M_{IJ}}{|M|^{2}}\mathcal{Z}^{*\, IJ}
+
\frac{M^{IJ}}{|M|^{2}}\mathcal{Z}_{IJ}
\right]
[\phi(\tau),\mathcal{Q}] d\tau\, ,
\end{equation}

\noindent
which leads to the component flow equation for the metric function

\begin{equation}
\frac{d}{d\tau} |M|^{-1}
=  -\tfrac{1}{\sqrt{2}}\Re{\rm e}
\left( \frac{M^{IJ}\mathcal{Z}_{IJ}}{|M|}
\right)\, .
\end{equation}

The final set of components of the flow equation ($N=2,3,4$) follows from

\begin{eqnarray}
{\textstyle\frac{1}{2}}\frac{M^{IJ}}{|M|^{2}} P_{iIJ} & = &  
-{\textstyle\frac{i}{2}} \frac{M^{IJ}}{|M|^{2}}
\langle\, d\mathcal{V}_{IJ} \mid \mathcal{V}_{i}\,  \rangle =
-{\textstyle\frac{i}{2}} 
\langle\, d\mathcal{V} \mid \mathcal{V}_{i}\,  \rangle
=  \langle\, d\mathcal{I} \mid \mathcal{V}_{i}\,  \rangle 
-{\textstyle\frac{i}{2}} 
\langle\, d\mathcal{V}^{*} \mid \mathcal{V}_{i}\,  \rangle \nonumber \\
& & \nonumber \\
& = & \tfrac{1}{\sqrt{2}}\mathcal{Z}_{i}[\phi(\tau),\mathcal{Q}]d\tau\, .
\end{eqnarray}

and takes the final form

\begin{displaymath}
P_{i\, KL}
\mathcal{J}^{K}{}_{I}\mathcal{J}^{L}{}_{J}  
=\sqrt{2}M_{IJ}\mathcal{Z}_{i}[\phi(\tau),\mathcal{Q}]d\tau\, .
\end{displaymath}

The equations for the critical (attractor) points of $N=8,d=4$ supergravity,
both supersymmetric and non-supersymmetric where given in
Ref.~\cite{Ferrara:2006em} and some flow equations for the $N=8$ theory based
on different Ansatze were given in Refs.~\cite{Andrianopoli:1997wi,Arcioni:1998mn}. It would be interesting to compare them with those
that are determined from the above flow equations, although more information
about the matrix of functions $M_{IJ}$ is necessary.


\section{Conclusions}
\label{sec-conclusions}

After considering two examples ($N=2,d=4,5$ supergravity) we have derived the
general flow equations for supersymmetric black holes in all $N\geq 2,d=4$
supergravities using a procedure that only uses basic properties of the scalar
manifolds of those theories and an Ansatz for certain combinations of the
scalar functions and some auxiliary functions ($M_{IJ}$) inspired in the
general form of the supersymmetric black-hole solutions. we have derived the
flow equations for $N\geq 2$ in a form which is manifestly duality-covariant.

The procedure used here may apply to many more solutions (supersymmetric of
not) which share the form of the Ansatz. For instance, it should apply to
non-supersymmetric extremal black holes and may also apply, for instance, to
cosmological solutions in which the coordinate $\tau$ is timelike. This
derivation, which depends on so few assumptions, may shed new light on the
reasons underlying the attractor mechanism in black holes and other
supergravity solutions.

The generalization of these derivations to higher dimensions should be
straightforward, using the formulation of Ref.~\cite{Andrianopoli:1996ve}.


\section*{Acknowledgments}

This work has been supported in part by the Spanish Ministry of Science and
Education grant FPA2009-07692, the Comunidad de Madrid grant HEPHACOS
S2009ESP-1473 and the Spanish Consolider-Ingenio 2010 program CPAN
CSD2007-00042.  The author wishes to thank M.M.~Fern\'andez for her permanent
support.

\appendix


\begin{thebibliography}{99}

\bibitem{Ferrara:1995ih}
S.~Ferrara, R.~Kallosh and A.~Strominger,
Phys.\ Rev.\ D {\bf 52} (1995) 5412
[\hepth{9508072}].

\bibitem{Strominger:1996kf}
A.~Strominger,
Phys.\ Lett.\ B {\bf 383} (1996) 39
[\hepth{9602111}].

\bibitem{Ferrara:1996dd}
S.~Ferrara and R.~Kallosh,
Phys.\ Rev.\ D {\bf 54} (1996) 1514
[\hepth{9602136}].

\bibitem{Ferrara:1996um}
S.~Ferrara and R.~Kallosh,
Phys.\ Rev.\ D {\bf 54} (1996) 1525
[\hepth{9603090}].

\bibitem{Ferrara:1997tw}
S.~Ferrara, G.~W.~Gibbons and R.~Kallosh,
Nucl.\ Phys.\ B {\bf 500} (1997) 75
[\hepth{9702103}].

\bibitem{Ferrara:2006em}
S.~Ferrara and R.~Kallosh,
Phys.\ Rev.\ D {\bf 73} (2006) 125005
[\hepth{0603247}].

\bibitem{Khuri:1995xq}
R.~R.~Khuri and T.~Ort\'{\i}n,
Phys.\ Lett.\ B {\bf 373} (1996) 56
[\hepth{9512178}].

\bibitem{Ortin:1996bz}
T.~Ort\'{\i}n,
Phys.\ Lett.\ B {\bf 422} (1998) 93
[\hepth{9612142}].

\bibitem{Dabholkar:1997rk}
A.~Dabholkar,
Phys.\ Lett.\ B {\bf 402} (1997) 53
[\hepth{9702050}].

\bibitem{Sen:2005wa}
A.~Sen,
JHEP {\bf 0509} (2005) 038
[\hepth{0506177}].

\bibitem{Goldstein:2005hq}
K.~Goldstein, N.~Iizuka, R.~P.~Jena and S.~P.~Trivedi,
Phys.\ Rev.\ D {\bf 72} (2005) 124021
[\hepth{0507096}].

\bibitem{Tripathy:2005qp}
P.~K.~Tripathy and S.~P.~Trivedi,
JHEP {\bf 0603} (2006) 022
[\hepth{0511117}].

\bibitem{Kallosh:2006bt}
R.~Kallosh, N.~Sivanandam and M.~Soroush,
JHEP {\bf 0603} (2006) 060
[\hepth{0602005}].

\bibitem{Bellorin:2006xr}
J.~Bellor\'{\i}n, P.~Meessen and T.~Ort\'{\i}n,
Nucl.\ Phys.\ B {\bf 762} (2007) 229
[\hepth{0606201}].

\bibitem{Ceresole:2007wx}
A.~Ceresole and G.~Dall'Agata,
JHEP {\bf 0703} (2007) 110
[\hepth{0702088}].

\bibitem{Andrianopoli:2007gt}
L.~Andrianopoli, R.~D'Auria, E.~Orazi, M.~Trigiante
JHEP {\bf 0711 } (2007)  032.
[\arxiv{0706.0712} [hep-th]].

\bibitem{Bossard:2009we}
G.~Bossard, Y.~Michel, B.~Pioline,
[\arxiv{0908.1742} [hep-th]].

\bibitem{Ceresole:2009vp}
A.~Ceresole, G.~Dall'Agata, S.~Ferrara, A.~Yeranyan,
Nucl.\ Phys.\  {\bf B832 } (2010)  358-381.
[\arxiv{0910.2697} [hep-th]].

\bibitem{Andrianopoli:2009je}
L.~Andrianopoli, R.~D'Auria, E.~Orazi, M.~Trigiante,
Nucl.\ Phys.\  {\bf B833 } (2010)  1-16.
[\arxiv{0905.3938} [hep-th]].

\bibitem{Ceresole:2010nm}
A.~Ceresole, S.~Ferrara, A.~Marrani,
Phys.\ Lett.\  {\bf B693 } (2010)  366-372.
[\arxiv{1006.2007} [hep-th]].

\bibitem{Gauntlett:2002nw}
J.P.~Gauntlett, J.B.~Gutowski, C.M.~Hull, S.~Pakis and H.S.~Reall,
Class.\ Quant.\ Grav.\  {\bf 20} (2003) 4587
[\hepth{0209114}].

\bibitem{Gauntlett:2004qy}
J.P.~Gauntlett and J.B.~Gutowski,
Phys.\ Rev.\ D {\bf 71} (2005) 045002
[\hepth{0408122}].



\bibitem{Lopes Cardoso:2007ky}
G.~Lopes Cardoso, A.~Ceresole, G.~Dall'Agata, J.~M.~Oberreuter, J.~Perz,
JHEP {\bf 0710 } (2007)  063.
[\arxiv{0706.3373} [hep-th]].

\bibitem{Mohaupt:2009iq}
T.~Mohaupt, K.~Waite,
JHEP {\bf 0910 } (2009)  058.
[\arxiv{0906.3451} [hep-th]].

\bibitem{Bellucci:2010aq}
S.~Bellucci, S.~Ferrara, A.~Shcherbakov, A.~Yeranyan,
[\arxiv{1010.3516} [hep-th]].

\bibitem{Sabra:1997kq}
W.~A.~Sabra,
Mod.\ Phys.\ Lett.\  {\bf A12 } (1997)  2585-2590.
[\hepth{9703101}].

\bibitem{Sabra:1997dh}
W.~A.~Sabra,
Nucl.\ Phys.\  {\bf B510 } (1998)  247-263.
[\hepth{9704147}].

\bibitem{Behrndt:1997ny}
K.~Behrndt, D.~Lust, W.~A.~Sabra,
Nucl.\ Phys.\  {\bf B510 } (1998)  264-288.
[\hepth{9705169}].

\bibitem{LopesCardoso:2000qm}
G.~Lopes Cardoso, B.~de Wit, J.~Kappeli, T.~Mohaupt,
JHEP {\bf 0012 } (2000)  019.
[\hepth{0009234}].

\bibitem{Meessen:2006tu}
P.~Meessen and T.~Ort\'{\i}n,
Nucl.\ Phys.\ B {\bf 749} (2006) 291
[\hepth{0603099}].

\bibitem{Andrianopoli:1996ve}
L.~Andrianopoli, R.~D'Auria and S.~Ferrara,
Int.\ J.\ Mod.\ Phys.\  A {\bf 13} (1998) 431
[\hepth{9612105}].

\bibitem{Meessen:2010fh}
P.~Meessen, T.~Ort\'{\i}n and  S.~Vaul\`a,
JHEP {\bf 11} (2010) 072.
[\arxiv{1006.0239} [hep-th]].

\bibitem{Huebscher:2006mr}
M.~H\"ubscher, P.~Meessen, T.~Ort\'{\i}n,
Nucl.\ Phys.\  {\bf B759 } (2006)  228-248.
[\hepth{0606281}].

\bibitem{Andrianopoli:1997wi}
L.~Andrianopoli, R.~D'Auria, S.~Ferrara, P.~Fre, M.~Trigiante,
Nucl.\ Phys.\  {\bf B509 } (1998)  463-518.
[\hepth{9707087}]
                
\bibitem{Arcioni:1998mn}
G.~Arcioni, A.~Ceresole, F.~Cordaro, R.~D'Auria, P.~Fre, L.~Gualtieri, M.~Trigiante,
Nucl.\ Phys.\  {\bf B542 } (1999)  273-307.
[\hepth{9807136}].

\bibitem{Andrianopoli:1997pn}
L.~Andrianopoli, R.~D'Auria, S.~Ferrara,
Phys.\ Lett.\  {\bf B403 } (1997)  12-19.
[\hepth{9703156}]

\bibitem{Ferrara:2007tu}
S.~Ferrara, A.~Marrani,
Phys.\ Lett.\  {\bf B652 } (2007)  111-117.
[\arxiv{0706.1667} [hep-th]].    

\bibitem{Andrianopoli:2010bj}
L.~Andrianopoli, R.~D'Auria, S.~Ferrara, M.~Trigiante,
JHEP {\bf 1008 } (2010)  126.
[\arxiv{1002.4340} [hep-th]].




\end{thebibliography}
\end{document}